\documentclass[twocolumn,floatfix,prl,superscriptaddress,showpacs,amssymb]{revtex4}

\usepackage{graphicx}
\usepackage{bm}

\expandafter\ifx\csname package@font\endcsname\relax\else
 \expandafter\expandafter
 \expandafter\usepackage
 \expandafter\expandafter
 \expandafter{\csname package@font\endcsname}%
\fi

\begin{document}

\title{Avalanche Dynamics in Wet Granular Materials}

\author{P. Tegzes}
\affiliation{Department of Physics and Materials Research Institute, %
Pennsylvania State University, University Park PA 16802}
\affiliation{Department of Biological Physics, E\"otv\"os Lor\'and University, %
1A P\'azm\'any stny., Budapest, Hungary 1117}
\author{T. Vicsek}
\affiliation{Department of Biological Physics, E\"otv\"os Lor\'and University, %
1A P\'azm\'any stny., Budapest, Hungary 1117}
\author{P. Schiffer}
\email{schiffer@phys.psu.edu}
\affiliation{Department of Physics and Materials Research Institute, %
Pennsylvania State University, University Park PA 16802}

\date{\today}

\begin{abstract}
We have studied the dynamics of avalanching wet granular media in a
rotating drum apparatus. Quantitative measurements of the flow
velocity and the granular flux during avalanches allow us to characterize
novel avalanche types unique to wet media.
We also explore the details of viscoplastic flow (observed at the highest
liquid contents) in which there are lasting contacts during flow,
leading to coherence across the entire sample.  This coherence leads to a
velocity independent flow depth at high rotation rates and novel robust 
pattern
formation in the granular surface.
\end{abstract}

\pacs{45.70.-n, 45.70.Ht, 45.70.Mg}

\maketitle

Avalanches and landslides \cite{Brabb89,Dikau96,mudflow} are among the
most dramatic of natural catastrophes, and they also provide an evocative
metaphor for a wide range of propagating breakdown phenomena from impact
ionization in semiconductors \cite{Clauss90} to magnetic vortex motion in
superconductors \cite{Field95}.  On the other hand, the existence of
avalanches, i.e. the sudden collapse of the system previously frozen into a
high energy state, is a fundamental manifestation of the metastable  nature 
of 
granular materials. Studies of avalanches \cite{Rajchenbach90,Liu91,dryava} 
and
surface flows \cite{dryflow} in granular media have largely
focused on dry grains. By wetting such media, however, one
introduces controllable adhesive forces between the grains
\cite{Schubert84,Craig97} which lead to qualitatively new behavior
\cite{Hornbaker97,Tegzes99,Halsey98,Bocquet98,Fraysse99,Nase01,Samadani00}. In our previous work \cite{Tegzes99} we identified three fundamental 
regimes
for the repose angle of wet granular materials as a function of the liquid
content. The granular regime at very low liquid contents is dominated by the
motion of individual grains; in the correlated regime corresponding to
intermediate liquid contents, a rough surface is formed by the flow of 
separated clumps; and the repose angle of very wet samples results from cohesive flow with viscoplastic properties.

Here we report investigations of the avalanche dynamics and flow properties
of wet granular materials, employing a rotating drum apparatus (a
cylindrical chamber partly filled with a granular medium and rotated around
a horizontal axis) \cite{Rajchenbach90,Liu91}.  At low rotation rates, the
medium remains at rest relative to the drum while its surface angle is
slowly increased by rotation, up to a critical angle ($\theta_{max}$) where
an avalanche occurs, thus decreasing the surface angle to the repose angle
($\theta_r$). The flow becomes continuous at high rotation rates, but the
 transition between avalanching and continuous flow is hysteretic in rotation rate in dry media \cite{Rajchenbach90,Caponeri95}. 

Previous studies of cohesive granular media
in a rotating drum \cite{Bocquet98,Fraysse99,Nase01} have focused on the
surface angles of the medium before and after avalanches. For example, Quintanilla
{\em et al.} recently performed a statistical analysis of avalanche size based on these angles, and showed that the
average clump size increased with cohesion \cite{Viturro01}. In our
measurements, we focus instead on characterizing the {\em dynamics of cohesive flow}.   We quantitatively investigate the flow dynamics during avalanches at different liquid contents by analyzing the time evolution of the averaged surface profile obtained from hundreds of avalanche events, and we also measure surface velocities during continuous flow.  In particular, we explore the nature of the
viscoplastic flow, which displays unique characteristics associated with
coherent motion over the entire granular surface.

We studied glass
spheres thoroughly mixed with small quantities of hydrocarbon oil
\cite{oilprop} varying between $\tau =$ $0.001\%$ and $5\%$ of the void
volume.  In this regime the flow of oil due to gravity can be neglected. 
The
inner diameter of our drum was $16.8$ cm, its width was $3.2$ cm, and the
granular filling was $30\%$. Measurements performed in a thinner ($2$ cm)
drum verified that wall effects do not modify the qualitative behavior.  We
studied two different sizes of beads ($d_1 = 0.5$\,mm$\pm 20\%$ and
$d_2 = 0.9$\,mm$\pm11\%$), and the results were qualitatively equivalent in 
the
two cases except where noted otherwise.  The experiment was recorded by a
CCD video camera interfaced to a computer which could analyze the
spatiotemporal evolution of the surface profile (height variations in the
axial direction were negligible). To determine the position of the surface,
we used background illumination and an algorithm based on local contrast
which gave us a resolution of $\delta t = 0.03$ seconds and $\delta x =
0.5$\,mm.

\begin{figure}
 \includegraphics[clip,width=\columnwidth]{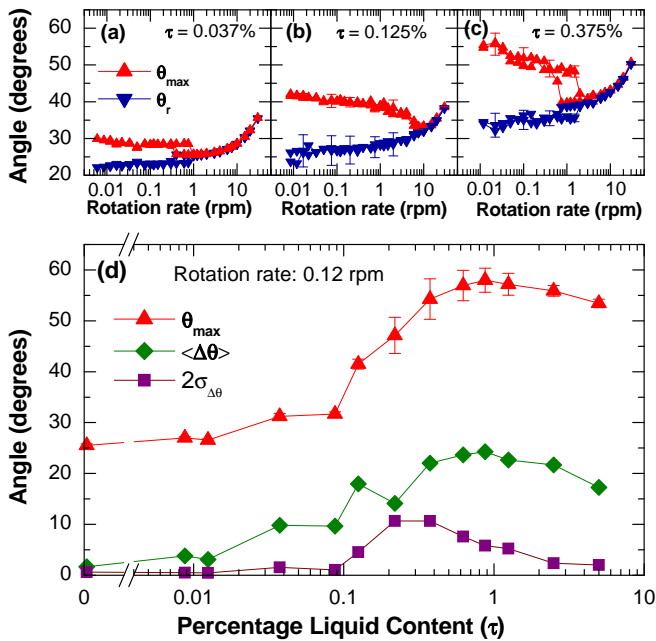}%
\caption{ (a) (b) (c) $\theta_{max}$ and $\theta_r$ as a
function of rotation rate for 3 different liquid contents. The error bars
represent the standard deviation of the observed angle distributions. (d)
The maximum angle, $\theta_{max}$, the average avalanche size,
$\langle\Delta\theta\rangle$, and the width of the avalanche size
distribution $2\sigma_{\Delta\theta}$ as a function of liquid content.
(The data were taken using small beads, $d=0.5$\,mm)} 
\label{Fig1}
\end{figure}

{\em Surface angles.}\ --- We first investigated the
avalanching -- continuous flow transition for various liquid contents. We
slowly increased and then decreased the rotation rate and measured 
the average surface angles,
$\theta_{max}$ and $\theta_r$ in the avalanching regime.
We consider the flow continuous when the medium never comes to rest with
respect to the drum, which coincides with
$\Delta\theta\equiv\theta_{max}-\theta_r<1^\circ$.
The results of typical runs are presented in Fig. \ref{Fig1} (a), (b) and
(c), where $\theta_{max}$ and $\theta_r$ are plotted as a function of the
rotation rate. 

The three  basic types of behavior observed
previously are also reflected in the present experiment. In the
granular regime at low liquid contents, (e.g. $\tau = 0.04\%$), the 
behavior
is qualitatively similar to the dry case, and we observe a clear hysteretic
transition between  continuous flow and avalanching (Fig. \ref{Fig1}
(a)). 
At a somewhat higher liquid content ($\tau = 0.125\%$, see Fig. \ref{Fig1}
(b)) the correlated behavior is marked by  a lack of hysteresis, and 
avalanching behavior is observed at a relatively high rotation rate 
compared to the other regimes. The continuous flow in this regime consists 
of a stream of separated clumps rather than the constant flux seen in the 
other two regimes.  At $\tau = 0.375\%$ (Fig. \ref{Fig1} (c)) hysteresis is 
again observed, reflecting the onset of the viscoplastic continuous flow, 
which is smooth and coherent over the entire sample. 

An interesting feature of the curves in Fig. \ref{Fig1} (a-c), is
that $\theta_{max}$ decreases with increasing rotation rate,
although at slow rotation (where the rotation of the drum during
the avalanche is negligible) the rotation rate influences the
avalanche process only through changes in the waiting time between
successive avalanches. We find that $\theta_{max}$ depends logarithmically 
on this
waiting time \cite{unpublished}, reminiscent of other granular aging 
phenomena 
\cite{Bocquet98,Losert00}.
The effect is more pronounced at higher liquid contents, and we
attribute it to the motion of liquid flowing towards the contact
points rather than condensation effects \cite{Bocquet98} since the
vapor pressure of the oil is low. 

If the rotation rate is sufficiently low, then discrete avalanches occur at all liquid contents. In the following paragraphs we analyze 
this avalanching regime. Fig. \ref{Fig1} (d) shows 
$\langle\theta_{max}\rangle$ (averaged over several hundred avalanches) as 
a function of liquid content. The distribution  is 
close to Gaussian, and the error bars in the figure indicate the standard 
deviation. The average avalanche size $\langle\Delta\theta\rangle= 
\langle\theta_{max}-\theta_r\rangle$ and the width of the avalanche size 
distribution $2\sigma_{\Delta\theta}=2(\langle\Delta\theta^2\rangle-
\langle\Delta\theta\rangle^2)^{1/2}$ are also plotted in Fig. \ref{Fig1} 
(d). We interpret these results below in the context of a more detailed 
investigation of the avalanche dynamics.

{\em  Flow Dynamics.} --- By using the rotating drum apparatus, we can 
obtain
information not just about the medium before and after the avalanche 
events,
 but we can also study the details of the grain motion {\em during avalanche events}.  In order to analyze the dynamics of avalanches we have obtained two dimensional space-time matrices,
$h(x,t)$, characterizing the sample surfaces throughout the avalanche
process which can then be analyzed to produce a variety of information 
about
the individual avalanches.  By taking derivatives of the
$h(x,t)$ data, we obtain the local angle ($\alpha (x,t)=\arctan [\partial 
_x
h(x,t)$]) and the local vertical velocity ($u(x,t)=\partial _t h(x,t)$) of
the surface profile. Furthermore, by then integrating the vertical velocity
(using the continuity equation and assuming constant density 
\cite{Douady98}), 
we also
obtain the local flux in the avalanche, i.e. $\phi (x,t) = \rho w
\int_{-D/2}^x \partial_t h(x',t) dx'$ (where $\rho$ is the grain density,
$w$ is the width of the drum, and $D$ is the drum diameter), which
represents the material flowing through a given vertical plane at position 
$x$. 

\begin{figure}
 \includegraphics[clip,width=\columnwidth]{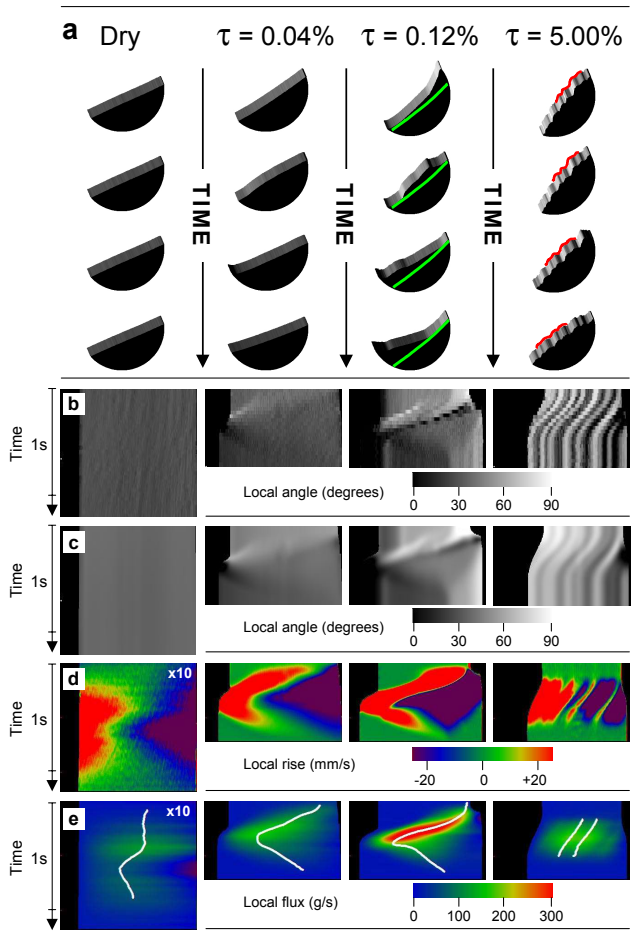}%
\caption{ Dynamics of avalanches of different types with grain
size $d=0.5$\,mm. (a) Snapshots (at $0.1 - 0.2$\,s time intervals) of
four single avalanches corresponding to different liquid contents.
The third dimension is used for a brightness-coded representation
of the local slope ($\alpha$).  The green lines indicate
approximate slip planes in the correlated regime, and the red line
shows the traveling quasi-periodic surface features in the
viscoplastic regime. (b) The local slope with the same brightness
coding, as a function of space and time.  A horizontal line
corresponds to a surface profile at a given instant, and time
propagates downwards (thus avalanches propagate leftwards and
down). The stripes indicate lasting surface features. (c) (d)
(e) The characteristic features of the avalanches averaged over
several hundred independent avalanches. The displayed quantities are: (c)
local surface angle, (d) the rate of change of local height, and
(e) the local grain flux (the white lines indicate the point of
maximum current). Note that the surface patterns at
the highest liquid contents are robust against averaging.
The standard deviations of the averaged data sets were typically $5-25$\,\% of the mean value.  In some cases, however, the noise 
originating from taking numerical derivatives leads to standard deviations 
comparable to the mean. For this reason averaging over several hundred 
independent events was absolutely necessary, resulting in standard errors ranging
from $0.2-6$\,\%.}
\label{Fig2}
\end{figure}

In Fig. \ref{Fig2} (a) we present snapshots of the progression of single
avalanches for several typical liquid contents. Fig. \ref{Fig2} (b) 
displays
$\alpha(x,t)$ as a function of space (horizontally) and time (downwards) 
for
the same individual avalanches. In Fig. \ref{Fig2} (c-e), we present the
average behavior of 300-500 avalanches at the same liquid contents, and 
show
similar graphs of the time evolution of
$\langle \alpha (x,t) \rangle$, $\langle u(x,t) \rangle$, and
$\langle \phi (x,t) \rangle$, where $\langle \rangle$ denotes
averaging over avalanches. By obtaining these quantitative measures of the
averaged properties, we can separate the robust characteristics of the
avalanche dynamics from the large fluctuations which are inherent in
avalanche processes.

In the avalanches among dry grains \cite{Rajchenbach90,Liu91,dryava},  the 
avalanches
consist of flow in a thin layer near the smooth surface and have a much 
longer 
duration and much smaller flux than in the wet media.
The size of the avalanches is $\Delta\theta\approx 
2^\circ\pm0.5^\circ$.
At low but nonzero liquid contents
(e.g. $\tau= 0.04\%$) $\theta_{max}$ and $\Delta\theta$ become larger due to the onset 
of intergrain cohesion. 
In this granular regime the avalanche is initiated by a
front traveling downhill (bright line in Fig. \ref{Fig2} (c)), but we can
also observe a second front traveling uphill (dark line).  The second front
is also apparent in the motion of the point of maximum flux,  Fig. 
\ref{Fig2} 
(e), and it corresponds to the  grains
reaching a solid barrier at the bottom  \cite{Makse97}.

At intermediate liquid contents (the correlated regime,
$0.1 \%\lesssim\tau \lesssim 0.5\%$), both  $\theta_{max}$ and $\Delta\theta$ increase substantially.  In this regime, the principal failure mechanism is a fracture along a
curved slip plane (approximated by the green lines in Fig. \ref{Fig2} (a)), 
analogous to the dynamics of a class of geological events known as 
``slides''
\cite{Dikau96}. In the figure there is a single slip plane,
which leads to large avalanches with a relatively narrow distribution.
However, at larger liquid contents the avalanches occur
through a succession of local slip events
followed by a large avalanche extending over the whole surface 
\cite{unpublished}. This leads to large fluctuations in the size of the 
avalanches (see Fig. \ref{Fig1} (d)) as observed  
in cohesive powders \cite{Viturro01}.
The medium becomes more cohesive with increasing liquid content,
and the  upward traveling front disappears for $\tau
\ge 0.2\%$ since the material moving as a connected block stops
coherently when it hits the bottom.  

At the highest liquid contents, $\tau \approx 0.5 - 1\%$, the onset of 
coherent viscoplastic flow is apparent through numerous qualitative changes 
in the dynamics of flow. The coherence of avalanche flow in this regime is
demonstrated by the parallel lines during the avalanche in Fig \ref{Fig2}
(b) and (c). This behavior is qualitatively
similar to a different class of geological event, called a "debris flow" or
"mudflow" \cite{Dikau96,mudflow}.  Since the whole surface moves coherently 
(rather than breaking
apart and evolving separately in different parts of the surface),
fluctuations are strongly suppressed \cite{Tegzes99}
and the avalanche size distribution is rather narrow (Fig. \ref{Fig1} (d)). 
The small decrease in $\theta_{max}$ is probably due to lubrication 
\cite{Tegzes99}.

A novel property of the viscoplastic avalanches is the
topology of the top surface which spontaneously forms a nearly
periodic pattern (seen in Fig.
\ref{Fig2} (a) and (b)). This surface structure is maintained essentially 
intact
during the avalanche, indicating that there are lasting contacts
in the flowing layer. Moreover, the same pattern in the surface features is 
reproduced at the end of each avalanche.  This is
demonstrated most clearly in Fig \ref{Fig2} (c), where the average of 347
avalanches of the $\tau = 5\%$ sample has the same features as the typical
individual avalanche shown in Fig. \ref{Fig2} (a) and (b).  The robust
nature of the surface structure of the wettest grains is in sharp contrast
to the other regimes where averaging completely smoothes out the smaller
surface features. We can understand this behavior as resulting from
coherence of the entire flow, which strongly reduces fluctuations in this
regime.  With minimal fluctuations, the final surface structure after each
avalanche is essentially the same, thus setting the same initial condition
for the next avalanche.  With the same initial conditions for each
avalanche, naturally the surface features are reproduced each time. Our
experiments with the larger ($d=0.9$\,mm) beads also revealed  pattern
formation, but with a smaller characteristic size corresponding to 8-10
grain diameters. The difference is probably due to the smaller ratio of the
cohesive forces to the gravitational forces on the grains.

\begin{figure}
 \includegraphics[clip,width=\columnwidth]{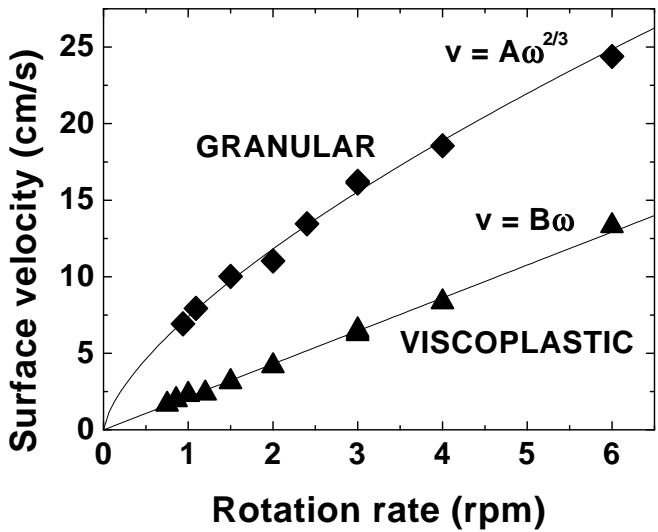}%
\caption{
The surface velocity $v$ during continuous flow as a function of rotation rate $\omega$ for dry grains and in the viscoplastic regime, $\tau=5\,$\%  (large beads,
$d=0.9$\,mm). The continuous lines represent power-low fits to the data.
The linear behavior for the
viscoplastic flow indicates that flow depth is independent of the rotation
rate.
}
\label{Fig3}
\end{figure}

 In addition to differences in the avalanching behavior, the viscoplastic 
regime also displays continuous flow which is rather different from that at lower liquid contents.  We compare the viscoplastic and granular continuous flows by measuring their surface velocities $v$ (by means of tracer particles) as a function of rotation rate in Fig. \ref{Fig3} (continuous flow is only observed in the correlated regime above $6$\,rpm, and we therefore do not include data from that regime here).  For these measurements we used 
the larger, ($d=0.9$\,mm) beads. 
The viscoplastic flow is slower than the granular flow by a factor of two,
and  the two curves correspond to rather different
functional forms. The curvature in the granular case suggests that  the 
flow 
depth increases with rotation rate  as  previously suggested
\cite{finitedepth}. The linear $v(\omega)$ function of the viscoplastic 
flow
suggests a {\em constant flow depth} which is independent of the flow rate.
This observation is consistent with the coherent nature of the flow: the 
flow 
depth is not determined by local mechanisms, but is fixed by
the geometry of the whole system \cite{dryflow}.

{\em Discussion.} --- While the flows we observe appear to have analogies
with geological events, it is important to note that  real geological 
materials 
usually consist of polydisperse
irregular particles often with very high ($\tau\approx100\%$) water 
contents
and that the scaling of our system to geological lengthscales is
non-trivial. Furthermore, avalanche studies in real soil have demonstrated
additional phenomena associated with soil saturation \cite{Iverson00}.
Interestingly we still recover some of the basic dynamical processes in our
model system, which constitutes an important step towards the description 
of 
qualitatively
different flow behaviors in the framework of a single model
\cite{Nase01,Gray01}.
The changes in the dynamical behavior with wetting are associated with the increasingly
coherent nature of the flow, i.e. the formation of coherently moving
clusters -- clumps -- due to the increased cohesion and viscous effects. 
Within a
cluster, local velocity fluctuations should be suppressed, and thus the
local granular temperature ($T = \langle v^2 \rangle - \langle v \rangle^2$) should approach zero, but
the clusters themselves both form and break apart during an avalanche
process in a finite container. An important theoretical question raised by our data is how a
length scale
describing the size of the clumps
may emerge from a granular flow model, and how such a length scale would
vary with the type of media, the nature of intergranular adhesion, and the
type of granular flow.

\begin{acknowledgments}

We gratefully acknowledge helpful discussions with J.
Banavar, A.-L. Barab\'asi, and Y. K. Tsui. We are also grateful for support
from the Petroleum Research Fund and the NASA Microgravity Fluid Physics
Program. P. T. and T. V. are grateful for the partial support from OTKA
Grant No. T033104.

\end{acknowledgments}


\begin{thebibliography}{1}
\bibitem{oilprop}{
The viscosity of the oil used was $0.27$ poise, and its surface tension was
$2\times 10^{-4}$  J/cm$^2$.
}
\bibitem{finitedepth}{
Naturally in our finite setup the flow depth cannot increase infinitely 
which 
may explain why our measured $v(\omega)$ deviates
from $\omega^{1/2}$ which would be expected from a linear velocity profile.
\cite{dryflow}
}
\end{thebibliography}
\end{document}